\documentclass[utf8]{FrontiersinHarvard} % for articles in journals using the Harvard Referencing Style (Author-Date), for Frontiers Reference Styles by Journal: https://zendesk.frontiersin.org/hc/en-us/articles/360017860337-Frontiers-Reference-Styles-by-Journal
\usepackage{url,hyperref,lineno,microtype,subcaption}
\usepackage[onehalfspacing]{setspace}

\usepackage{graphicx}   % for \rotatebox
\usepackage{array}      % column formatting
\usepackage{multirow}   % multirow cells (optional here)
\usepackage{colortbl,xcolor} % cell colors
\usepackage{makecell}   % better cell handling
\usepackage{adjustbox}  % to fit table to width

\usepackage{graphicx}
\usepackage{subcaption}
\usepackage{caption}

\usepackage{listings}
\usepackage{xcolor}
\def\keyFont{\fontsize{8}{11}\helveticabold }
\def\firstAuthorLast{Talburt {et~al.}} %use et al only if is more than 1 author
\def\Authors{John R. Talburt$^{1}$, Muzakkiruddin Ahmed Mohammed$^{1}$, Mert Can Cakmak $^{1}$, Onais Khan Mohammed$^{1}$, Mahboob Khan Mohammed$^{1}$, Khizer Syed$^{1}$, Leon Claasssens$^{2}$}
\def\Address{$^{1}$ Center for Advanced Research in Entity Resolution and Information Quality (ERIQ) \\ University of Arkansas - Little Rock \\
$^{2}$ Pilog Group }
\def\corrAuthor{Mert Can Cakmak}

\def\corrEmail{mccakmak@ualr.edu}

\begin{document}
\onecolumn
\firstpage{1}

\title[Case Count Metric for Comparative Analysis of Entity Resolution Results]{Case Count Metric for Comparative Analysis of Entity Resolution Results} 

\author[\firstAuthorLast ]{\Authors} %This field will be automatically populated
\address{} %This field will be automatically populated
\correspondance{} %This field will be automatically populated

\extraAuth{}% If there are more than 1 corresponding author, comment this line and uncomment the next one.
%\extraAuth{corresponding Author2 \\ Laboratory X2, Institute X2, Department X2, Organization X2, Street X2, City X2 , State XX2 (only USA, Canada and Australia), Zip Code2, X2 Country X2, email2@uni2.edu}

\maketitle

\begin{abstract}

This paper describes a new process and software system, the Case Count Metric System (CCMS), for systematically comparing and analyzing the outcomes of two different ER clustering processes acting on the same dataset when the true linking (labeling) is not known. The CCMS produces a set of counts that describe how the clusters produced by the first process are transformed by the second process based on four possible transformation scenarios. The transformations are that a cluster formed in the first process either remains unchanged, merges into a large cluster, is partitioned into smaller clusters, or otherwise overlaps with multiple clusters formed in the second process. The CCMS produces a count for each of these cases accounting for every cluster formed in the first process. In addition, when run in analysis mode, the CCMS program can assist the user in evaluating these changes by displaying the details for all changes or only for certain types of changes. The paper includes a detailed description of the CCMS process and program and examples of how the CCMS has been applied in university and industry research.

\tiny
 \keyFont{ \section{Keywords:} Entity Resolution, Record Linkage, Evaluation Metrics, Comparative Analysis, Clustering Evaluation, Data Quality} %All article types: you may provide up to 8 keywords; at least 5 are mandatory.
\end{abstract}

\section{Introduction}

Entity Resolution (ER) is crucial for effective data management, particularly when dealing with large and complex datasets. ER is usually defined as the process of determining if two information system references (sometimes called “mentions” or “observations”) to real world objects are referring to the same, or to different, objects \citep{binette2022almost} In the case that two references are to the same object, the references are said to be equivalent \citep{talburt2011entity}. 

ER is usually implemented in one of two forms, cluster ER or binary ER \citep{binette2022almost}. The most basic and oldest ER process is binary ER \citep{fellegi1969theory} initially developed to reconcile census of a given population taken at different times. In binary ER, the input comprises two distinct datasets with the goal of identifying and linking equivalent references between the two datasets. Often the assumption is that there are no equivalent references within each dataset and as a result, a reference in the first dataset can link to at most one reference in the second dataset resulting in one-to-one binary ER. If this condition is relaxed, the linking can be one-to-many between the datasets.

However, the focus of this paper is on cluster ER. In cluster ER, in the input is a single dataset and the objective is to partition the dataset into non-intersecting subsets (clusters) where each cluster represents the entirety of references in the dataset referencing a particular entity. A formal mathematical definition of the entity resolution of a given set of references was developed at the Stanford InfoLab \citep{benjelloun2009swoosh}. In cluster ER, all the references in the same cluster are equivalent, and references in different clusters are not equivalent. In terms of a graph where the nodes are references and links are edges, a cluster is a maximally connected component of the graph. 

While binary ER relies entirely on direct linking of references, i.e., matching references that meet a specific level of similarity, cluster ER includes two additional indirect linking processes. The first indirect linking process is transitive closure. If one accepts the unique reference assumption for the input dataset \citep{talburt2015entity} then reference equivalence is a true equivalence relation in the mathematical sense, i.e., it is a binary set relation that is idempotent, symmetric, and transitive \citep{rotman2010advanced}. This allows non-matching references to be indirectly linked as equivalent through a chain of direct matching links. For example, if Reference A and Reference B are considered equivalent because A matches B, and if Reference B and C are considered equivalent because B matches C, then by transitive closure it follows that A and C are equivalent even if A and C do not match. The addition of indirect linking sometimes leads to confusion in the evaluation of the binary ER results versus cluster ER results. In binary ER, only direct matches are evaluated as either true or false positive links. However, in cluster ER, all pairs of references in the same cluster (both directly and indirectly linked) are considered linked pairs and subject to evaluation as either true or false positive links \citep{ye2018effect}.

The second indirect linking process in cluster ER is by patterns. A common indirect linking pattern is the household move pattern often observed in demographic data where members of the same household are observed residing at two or more addresses \citep{fu2014graph}. For example, the four-way pattern that person named John Doe is found residing at both Oak Street and Elm Street, and that another person named Mary Doe is also found residing at the same two addresses provides evidence (an increased probability) that these could be the same persons. While the two John Doe references and the two Mary Doe references might not rise to level of a matching pair, they still might be indirectly linked based on this household move pattern. Especially if there is additional evidence such as age similarity, low frequency names, more than two household members, or additional addresses for the same household members.

\section{Problem Statement}

Master data management (MDM) is a widely used data curation process employed by most large organizations to facilitate accurate data integration at scale. Cluster ER is a foundational process in all MDM systems. From an academic perspective, the typical way of evaluating cluster ER is to use ER metrics such as the F-measure, Precision, and Recall or weighted versions of these measures \citep{christen2023review}. However, these metrics rely upon having a truth set that allows every pair of input references to be definitively judged as either a true positive, false positive, true negative, or false negative, thus providing an accurate measure of the overall outcome. When a truth set is available, it is easy to compare two different cluster ER outcomes and judge that one outcome is better than the other. However, for most real-world applications, the true linking for any sizeable sample of references is unknown. While there are stratified sampling techniques to approximate the overall Precision and Recall for the clustering of large datasets \citep{pullen2017system} \citep{penning2016inferred} these methods are both labor-intensive and time-consuming.

Besides precision, recall, and F-measure metrics, there are other methods of comparing cluster ER outcomes. The confusion matrix showing the false positives \& false negative outcomes is crucial for comprehending the kinds of mistakes, false positive and false negative links, the system makes. Its application is advised by Herzog, Scheuren, and Winkler \citep{herzog2007data} as a thorough way to evaluate and contrast the accuracy of ER systems visually is essentially the same as the precision and recall measures. However, this method also requires the availability of a linking truth set.

When assessing and contrasting how well they perform of ER systems under changing decision criteria, the Receiver Operating Characteristic (ROC) curve and the Area Under the Curve (AUC) serve as vital tools. These could be especially helpful for modifying an ER system's sensitivities to satisfy certain operational needs, including reducing false negatives in identifying fraud software or increasing accuracy in focused marketing efforts \citep{hand2018note}. Again, a linking truth set is required for this method.

\section{Related Work}

Some alternative methods for evaluating ER outcomes without a truth set are in use in the cluster ER community. One of these methods is benchmarking. Benchmarking uses a standard dataset with established and acceptable results. The benchmark dataset offers a controlled setting for evaluating each system's advantages and disadvantages in relation to various information issues, such as different data quality levels, changes to linking logic, or comparing competing vendor tools. Benchmarking provides insights into the scalability and flexibility of various systems by demonstrating how they operate beneath comparable circumstances \citep{christen2012data}.  

Another approach is to use simulated or synthetically generated reference data. This is most often done for personal demographic data items which are considered sensitive such as name and address information. For example, synthetically generated occupancy information |\citep{talburt2009sog} and the PseudoPeople project \citep{haddock2024simulated} for census data. Large language models also show promise to generate synthetic data \citep{meng2022generating}. While this method can solve the truth set problem, its usefulness depends upon the degree to which the simulated references emulate real-world references.

Case studies can be of use in understanding the practical applicability and performance of ER systems. In addition to providing empirical support for theoretical models, real-world case studies highlight pragmatic issues such as integrating with pre-existing IT systems, managing real-time data streams, and adjusting to changing data environments. However, most of these studies are from government or non-profit organizations \citep{winkler2006data}. Commercial entities are reluctant to publish such results for reasons of competitiveness and confidentiality. 

In practice, most organizations, especially commercial entities, rely on experience and analysis over time to establish that their system has achieved nominal stability, reasonable accuracy, and user-acceptable outcomes. Because these are production systems, changes to these systems are only made carefully and incrementally. The impact of a change (intervention) is assessed on benchmark datasets to observe its net effect, i.e., by comparing whether the differences in clustering resulting from the changes are better or worse overall than the clusters in the unchanged, baseline system. However, these cluster change assessments are largely ad hoc and not systematic in nature.

To help with this problem, the TWI metric \citep{talburt2008algebraic} was introduced to measure the degree of difference between two cluster outcomes, A and B, for the same underlying dataset. The TWI is defined as

\[
TWI = \frac{\sqrt{|A| \cdot |B|}}{|V|}
\]

Where $|A|$ represents the number of clusters in outcome $A$, and $|B|$ the number of clusters in outcome $B$, and $|V|$ represents the number of non-empty intersections (overlaps) between the clusters in $A$ and $B$. The TWI is normalized to the interval $[0, 1]$ so that the value $1$ is only achievable if $A$ and $B$ are identical outcomes. While the TWI can never become $0$, the worst case is when $|A| = 1$ and $|B| = N$ (where $N$ is the number of references in the underlying dataset). In this case, $|V| = N$, and the TWI is the reciprocal of the square root of $N$, a smaller and smaller number as $N$ increases. While TWI does provide a relative measure of clustering difference without relying on the knowledge of the true linking, it does not provide any guidance on what these differences are.

\section{Methodology}

Recognizing the limitations inherent in existing metrics and methods, this paper proposes a novel method, the Cluster Count Metric System (CCMS), specifically designed to provide a more granular analysis of the dynamics of cluster ER changes without recourse to a truth set. 

\subsection{The Cluster Count Metric System (CCMS)}

Let $ER_1$ represent the set of clusters generated by the first (baseline) ER process acting on a given set of entity references $R$. Let $ER_2$ represent the set of clusters generated by the second ER process acting on $R$. $ER_1$ and $ER_2$ are the inputs to CCMS, and the output produced by CCMS is a series of counts.

The first set of counts produced by CCMS provides an overall profile of $R$, $ER_1$, and $ER_2$. These include:

\begin{itemize}
    \item $RC =$ Count of references in $R$
    \item $CC_1 =$ Count of clusters generated by $ER_1$
    \item $SC_1 =$ Count of singleton clusters generated by $ER_1$ (singleton clusters are clusters containing only one reference)
    \item $CC_2 =$ Count of clusters generated by $ER_2$
    \item $SC_2 =$ Count of singleton clusters generated by $ER_2$
\end{itemize}

However, the primary CCMS output is a second set of four counts that describe how each cluster formed by $ER_1$ is transformed by $ER_2$. CCMS recognizes four distinct, mutually exclusive transformations of a cluster in $ER_1$ to one or more clusters in $ER_2$. For a given cluster $A_{ER1}$, these counts are described as follows:

\textbf{Unchanged Count (UC):} A count of cases where the $ER_1$ cluster is identical to an $ER_2$ cluster. In other words, through their linking decisions, both $ER_1$ and $ER_2$ have formed the same cluster of references. 

\[
A = B, \quad \text{where } B \in ER_2
\]

\textbf{Merged Count (MC):} A count of cases where the entire $ER_1$ cluster is a proper subset of an $ER_2$ cluster. The $ER_1$ cluster becomes part of (merges into) a larger $ER_2$ cluster. In this case, all the references in the $ER_1$ cluster were also linked by the $ER_2$ process, but the $ER_2$ process made additional links to these references that were not made by the $ER_1$ process.

\[
A \subset B \ \text{ and }\ A \neq B, \quad \text{where } B \in ER_2
\]

\textbf{Partitioned Count (PC):} A count of cases where the $ER_1$ cluster is decomposed into multiple $ER_2$ clusters. In this case, the $ER_2$ process partitioned the $ER_1$ cluster into smaller clusters without adding any new references. Each of the $ER_2$ clusters having a non-empty intersection with the $ER_1$ cluster is a proper subset of the $ER_1$ cluster.

\[
A = \bigcup_{i=1}^{n} B_i, \quad \text{where } B_i \in ER_2 \ \text{ and }\ n>1
\]

\textbf{Overlapping Count (OC):} A count of the cases where the $ER_1$ cluster has a non-empty intersection with two or more $ER_2$ clusters and at least one intersecting $ER_2$ is not a subset of the $ER_1$ cluster. In other words, the $ER_1$ cluster is a proper subset of the union of the $ER_2$ clusters having a non-empty intersection with the $ER_1$ cluster. Unlike the partitioned case, this indicates an overlap where the $ER_1$ cluster is divided, but the resulting $ER_2$ clusters include more references than the original, suggesting a more complex reorganization.

\[
A \subset \bigcup_{i=1}^{n} B_i,\quad
A \neq \bigcup_{i=1}^{n} B_i, \quad
\text{where } B_i \in ER_2 \ \text{ and }\ n>1
\]

Because these cases are mutually exclusive and exhaustive, it also follows that these four counts sum to the total number of $ER_1$ clusters, i.e.,

\[
CC_1 = UC + MC + PC + OC
\]

It should be noted that the CCMS process is not symmetric. If the $ER_2$ clusters are considered as the baseline clusters, and the $ER_1$ clusters are considered as the transformed clusters ($ER_2$ vs $ER_1$), then the counts will be different. Although it is clear the unchanged count $UC$ will be the same for both $ER_1$-to-$ER_2$ and $ER_2$-to-$ER_1$, but if $CC_1 \neq CC_2$, then at least one of the other counts will be different between $ER_1$-to-$ER_2$ and $ER_2$-to-$ER_1$.

\begin{figure}[h!]
    \centering
    \includegraphics[width=0.54\linewidth]{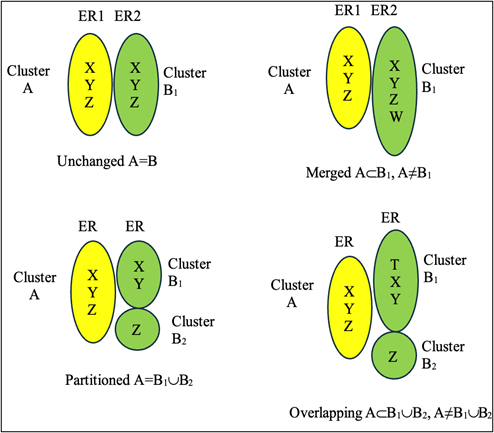}
    \caption{Graphical Representation of Cluster Cases}
    \label{cluster}
\end{figure}

A graphical representation of $ER_1$-to-$ER_2$ transformations are shown in Figure \ref{cluster} where X, Y, Z, and W represent references in $R$.

Table~\ref{tab:cluster_counts} shows a simple example of $ER_1$ vs $ER_2$ where $RC = 16$, $CC_1 = 7$, and $CC_2 = 8$ for a 16-reference dataset illustrating all four cases. Note that the rows are arranged in order by the $ER_1$ cluster identifier. The $ER_1$ clusters are highlighted with alternate shading.

\begin{table}[h]
\centering
\setlength{\tabcolsep}{6pt}
\renewcommand\arraystretch{1.15}
\begin{adjustbox}{max width=\linewidth}
\begin{tabular}{|c|c|c|c|c|c|c|}
\hline
\multicolumn{7}{|c|}{\textbf{Cluster comparison $ER_1$ to $ER_2$}} \\ \hline
\textbf{Reference IDs} & \textbf{$ER_1$ Cluster IDs} & \textbf{$ER_2$ Cluster IDs} & \textbf{Unchanged} & \textbf{Merged} & \textbf{Partitioned} & \textbf{Overlapping} \\ \hline
1  & a & x & 1 &  &  &  \\ \hline
\rowcolor{gray!50}
2  & b & y & 1  &  &  &  \\
\rowcolor{gray!50}
3  & b & y &  &  &  &  \\ \hline
4  & c & z &  &  &  &  \\ 
5  & c & z &  & 1  &  &  \\ 
6  & c & z &  &  &  &  \\ \hline
\rowcolor{gray!50}
7  & d & z &  & 1 &  &  \\ \hline
8  & e & w &  &  &  &  \\ 
9  & e & w &  &  & 1 &  \\ 
10 & e & t &  &  &  &  \\ \hline
\rowcolor{gray!50}
11 & f & u &  &  &  &  \\ 
\rowcolor{gray!50}
12 & f & u &  &  &  & 1  \\ 
\rowcolor{gray!50}
13 & f & v &  &  &  &  \\ \hline
14 & g & u &  &  &  &  \\ 
15 & g & v &  &  &  & 1 \\ 
16 & g & s &  &  &  &  \\ \hline
\textbf{Totals} & \textbf{7} & \textbf{8} & \textbf{2} & \textbf{2} & \textbf{1} & \textbf{2} \\ \hline
\multicolumn{7}{|l|}{\textbf{Total $ER_1$ Clusters (}$CC_1$\textbf{)} = 7 \{a, b, c, d, e, f, g\} \textbf{, Total $ER_2$ Clusters (}$CC_2$\textbf{)} = 8 \{x, y, z, s, t, u, v, w\}  } \\ \hline
\multicolumn{7}{|l|}{\textbf{$ER_1$ Singleton Clusters (}$SC_1$\textbf{)} = 2 \{a, d\}, \textbf{$ER_2$ Singleton Clusters (}$SC_2$\textbf{)} = 3 \{x, t, s\}} \\ \hline
\multicolumn{7}{|l|}{\textbf{Non-empty intersections between $ER_1$ and $ER_2$ clusters = 11}} \\ \hline
\multicolumn{7}{|l|}{$TWI = \dfrac{\sqrt{7 \times 8}}{11} = 0.68$} \\ \hline
\end{tabular}
\end{adjustbox}
\caption{\textbf{Example Cluster Counts}}
\label{tab:cluster_counts}
\end{table}

\subsection{Python Implementation of CCMS}

The Case Count Metric System (CCMS) has been implemented as a Python-based web application using Flask, enabling users to analyze, compare, and visualize the outcomes of Entity Resolution (ER) clustering processes. This implementation provides an interactive platform for cluster analysis when a truth set is unavailable. Below, we detail the design and functionality of this implementation.

\subsubsection{Design and Architecture}

\begin{figure}[htbp]
    \centering
    \includegraphics[width=0.6\linewidth]{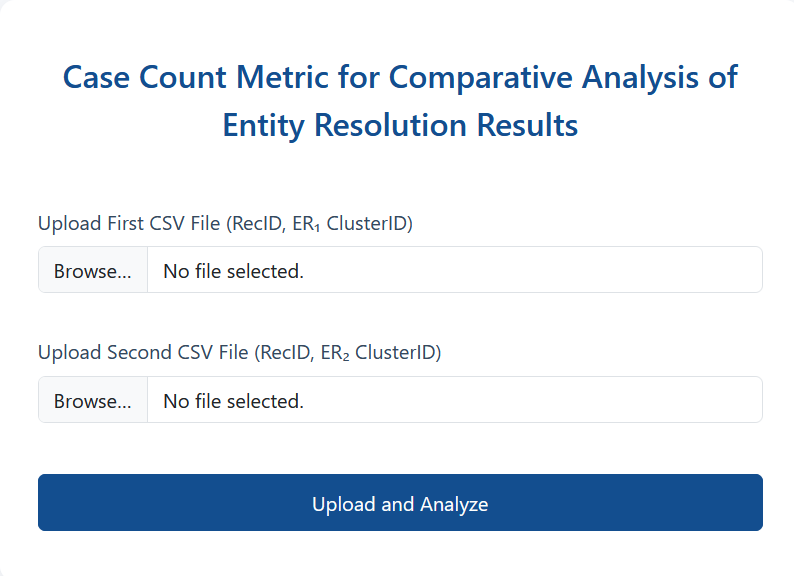}
    \caption{\textbf{User Input Interface for the Case Count Metric System.}
    This figure illustrates the application’s upload interface, where users provide two CSV files corresponding to ER$_1$ and ER$_2$ cluster outputs. The interface allows convenient data submission for comparative entity resolution analysis, facilitating visualization and metric computation.}
    \label{fig:input_interface}
\end{figure}

The application framework is developed using Flask to manage server-side operations such as file uploads, cluster analysis, and result rendering as it shown in Figure \ref{fig:input_interface}. It is integrated with a responsive HTML interface built using Bootstrap for enhanced usability and DataTables for efficient display of input data. The system accepts two CSV input files containing the \textit{RecID}, \textit{ER1 ClusterID}, and \textit{ER2 ClusterID} from two Entity Resolution (ER) systems. The output consists of a detailed cluster transformation analysis that includes textual summaries of cluster cases (Unchanged, Merged, Partitioned, and Overlapping), visualizations such as bar and pie charts illustrating the distribution of these cases, and identification of singleton clusters in both $ER_1$ and $ER_2$.

\subsubsection{Key Functional Components}

\textbf{Cluster Analysis and Classification}

Clusters from $ER_1$ and $ER_2$ are analyzed to classify transformations into four distinct cases:

\begin{itemize}
    \item \textbf{Unchanged:} Clusters that are identical in both $ER_1$ and $ER_2$.
    \item \textbf{Merged:} Clusters from $ER_1$ that are subsets of larger clusters in $ER_2$.
    \item \textbf{Partitioned:} Clusters from $ER_1$ that are split into multiple smaller clusters in $ER_2$.
    \item \textbf{Overlapping:} Clusters from $ER_1$ that overlap with multiple clusters in $ER_2$, forming complex relationships.
\end{itemize}

\textbf{Example Logic for Case Determination}

The following Python code illustrates the logical conditions used to identify \textit{identical} and \textit{partitioned} cluster cases:

\begin{lstlisting}[language=Python]
def is_identical(cluster):
    return len(cluster['er2_clusters']) == 1 and \
           cluster['size_er1'] == len(cluster['er2_references'])

def is_partitioned(cluster):
    return len(cluster['er2_clusters']) > 1 and \
           cluster['size_er1'] == len(cluster['er2_references'])
\end{lstlisting}

\textbf{Visualizations}

ECharts is utilized to generate interactive bar and pie charts that effectively illustrate the outcomes of cluster transformations. These visualizations provide a clear comparative overview of the following aspects:
\begin{itemize}
    \item Case count distributions across transformation types.
    \item Proportions of total clusters between $ER_1$ and $ER_2$.
\end{itemize}

\begin{figure}[htbp]
    \centering
    \includegraphics[width=0.6\linewidth]{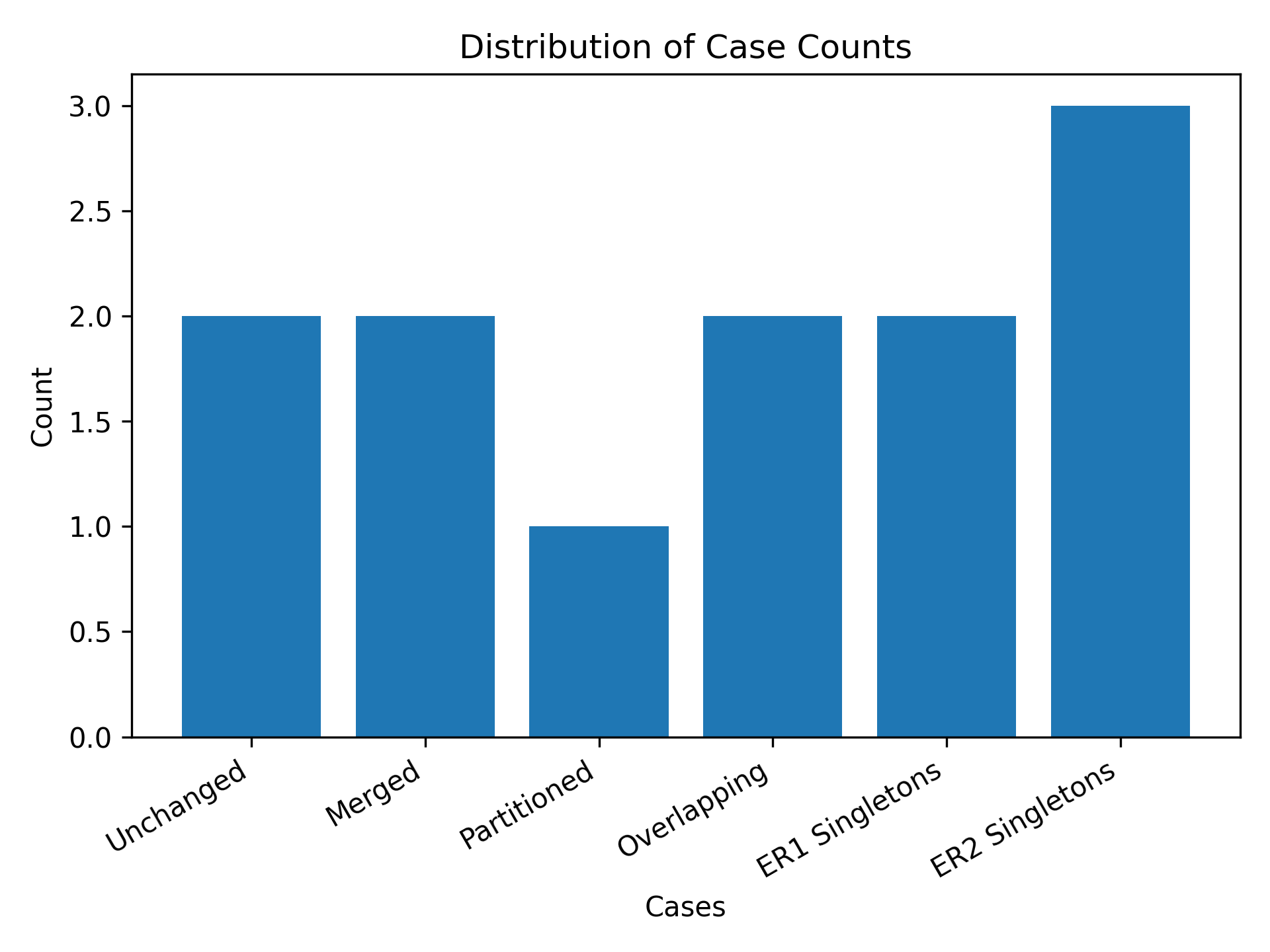}\par\vspace{-0.5em}
    \includegraphics[width=0.6\linewidth]{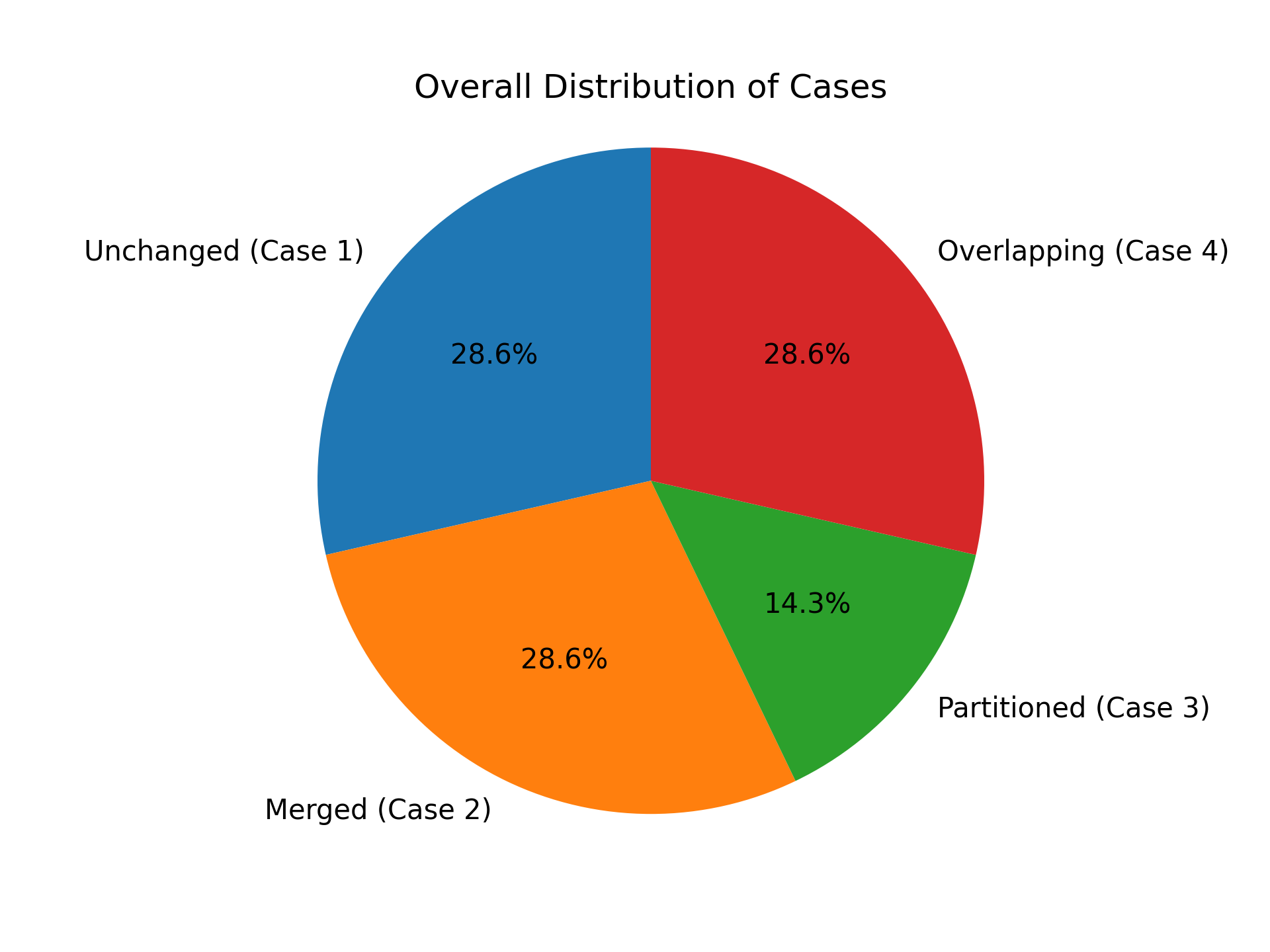}\par\vspace{-2em}
    \includegraphics[width=0.6\linewidth]{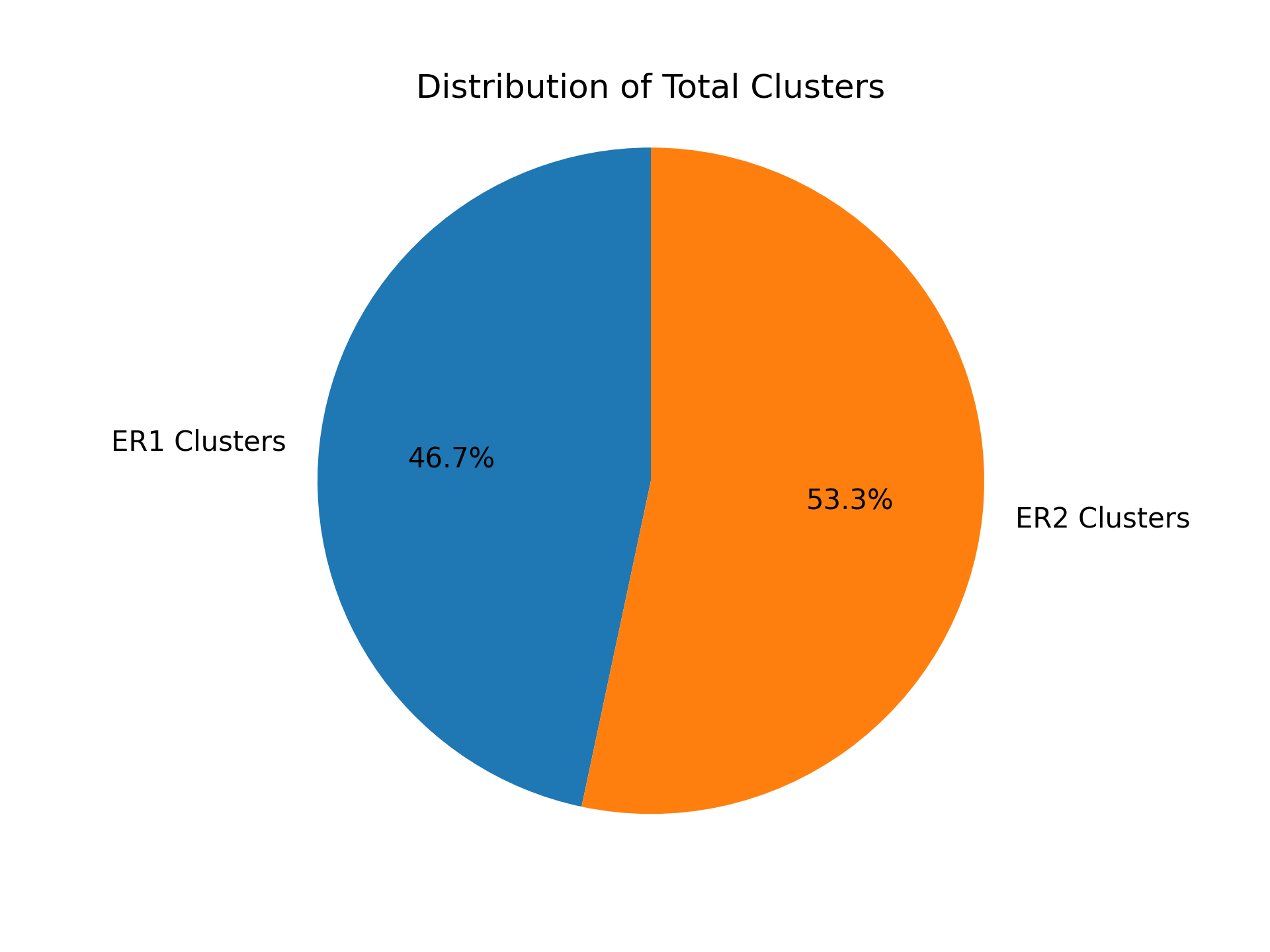}
    \caption{\textbf{Comparative Visualizations of Case Counts and Cluster Proportions.}
    Top: bar chart of case counts. Middle: pie chart of case proportions. Bottom: ER$_1$ vs.\ ER$_2$ cluster totals.}
    \label{fig:case_visualizations_stacked}
\end{figure}

The interactive charts enable researchers to explore the relationships between clustering outcomes, assess transformation frequencies, and visually identify outlier behavior in entity resolution processes. An example has shown in Figure \ref{fig:case_visualizations_stacked}. 

% No subcaption/caption packages needed for this

\textbf{Singleton Detection}

Singleton detection is implemented to identify isolated clusters in both $ER_1$ and $ER_2$, which often represent unique records or unlinked entities. The following function provides an example implementation for identifying singleton clusters:

\begin{lstlisting}[language=Python, caption={Example Python function for singleton cluster detection}]
def determine_singletons(df, clusters):
    er1_singletons = {
        cluster: refs for cluster, refs in clusters.items()
        if len(refs['er2_references']) == 1
    }
    return er1_singletons
\end{lstlisting}

Singleton clusters are subsequently summarized to aid in understanding the granularity and completeness of entity resolution results.

\textbf{Summary Report Generation}

The summary report consolidates all computed metrics, providing a textual overview of the clustering outcomes between $ER_1$ and $ER_2$. The report includes case counts, singleton cluster identification, and total cluster summaries for both directions of comparison.

\textbf{Example Output (from the application):}

\begin{verbatim}
Detailed Summary Report
ER1 as primary and ER2 as secondary:
Unchanged (Case 1): 2
Merged (Case 2): 2
Partitioned (Case 3): 1
Overlapping (Case 4): 2
ER1 clusters: 7

ER2 as primary and ER1 as secondary:
Unchanged (Case 5): 2
Merged (Case 6): 3
Partitioned (Case 7): 1
Overlapping (Case 8): 2
ER2 clusters: 8

Total clusters: 15

Singletons:
ER1 Singletons: 2
ER2 Singletons: 3
\end{verbatim}

\textbf{Example Use Case:}

\textit{Input:} Two sample CSV files containing the following data fields:
\begin{itemize}
    \item \texttt{RecID, ER1 ClusterID}
    \item \texttt{RecID, ER2 ClusterID}
\end{itemize}

\textit{Output:}
\begin{itemize}
    \item Visualizations (bar and pie charts) showing distribution of Unchanged, Merged, Partitioned, and Overlapping cluster cases.
    \item A detailed textual report summarizing case counts, singleton clusters, and overall cluster transformations.
\end{itemize}

The Python-based implementation of the Cluster Count Metric System (CCMS) provides a practical, user-friendly platform for comprehensive analysis and visualization of cluster differences between two entity resolution processes. This implementation bridges the gap between theoretical ER metrics and their application to real-world datasets, offering interpretable insights even in the absence of a ground truth set.

\section{Example Applications of CCMS}

The examples here give two different uses of the CCMS. The first example uses a synthetic demographic dataset where the truth set is known. In this application, the goal is to understand how sensitive a cluster ER system is to changes in certain continuous valued control parameters. In this case, both the input dataset and the ER system are held fixed while the interventions are a change in parameter value.

In the second example, the truth set is not known, and the intervention is a change in systems. In the second, example, the goal is to determine which system is giving better results through the analysis of cluster changes. 

\subsection{University Research: ER Parameter Sensitivity Analysis}

The first example relates to effect and sensitivity of the parameters controlling an unsupervised cluster ER system called the Data Washing Machine (DWM) \citep{talburt2020iterative, al2021model, anderson2023optimal}. The DWM is controlled by 29 parameters. While many of these deal with basic issues such as input location and output formatting, 14 of the parameters regulate the DWM’s unsupervised ER functions including blocking, linking, and cluster evaluation.

To adjust the parameter values whether manually or automatically, it is helpful to understand the impact on the DWM’s final clustering and the sensitivity to value changes. Here is where the CCMS can be helpful. As a simple example, consider the DWM parameter “mu”. This parameter sets the match level threshold and must be set to a value in the [0, 1] interval.

Table \ref{table2} shows the CCMS output where the input dataset (S8P) comprises 1,000 synthetic references generated by SOG \citep{talburt2009sog}. In this case, the ER1 output was produced where the ER parameters were set at what was believed to be an optimal setting including a mu value of 0.67. With this parameter configuration, ER1 produces 255 clusters. While the ER1 parameters are held fixed, each row of Table 3 represents a DWM clustering (ER2) using the same parameters as ER1 except for a change in the mu value. In this example, the baseline value of mu value is 0.67, the intervention was incrementing and decrementing mu from the baseline value in increments of 0.10.

\begin{table}[ht]
\centering
\setlength{\tabcolsep}{6pt}
\renewcommand\arraystretch{1.2}
\begin{adjustbox}{max width=\linewidth}
\begin{tabular}{|c|c|c|c|c|c|c|c|c|c|}
\hline
\textbf{Run} & \textbf{$\mu$} & \textbf{CC1} & \textbf{SC1} & \textbf{UC} & \textbf{MC} & \textbf{PC} & \textbf{OC} & \textbf{CC2} & \textbf{SC2} \\ \hline
1 & 0.17 & 255 & 94 & 8 & 247 & 0 & 0 & 9 & 7 \\ \hline
2 & 0.27 & 255 & 94 & 8 & 247 & 0 & 0 & 9 & 7 \\ \hline
3 & 0.37 & 255 & 94 & 10 & 245 & 0 & 0 & 11 & 8 \\ \hline
4 & 0.47 & 255 & 94 & 26 & 229 & 0 & 0 & 36 & 13 \\ \hline
5 & 0.57 & 255 & 94 & 76 & 179 & 0 & 0 & 108 & 35 \\ \hline
\rowcolor{yellow!50}
6 & 0.67 & 255 & 94 & 255 & 0 & 0 & 0 & 255 & 94 \\ \hline
7 & 0.77 & 255 & 94 & 150 & 0 & 105 & 0 & 451 & 244 \\ \hline
8 & 0.87 & 255 & 94 & 119 & 0 & 136 & 0 & 629 & 408 \\ \hline
9 & 0.97 & 255 & 94 & 126 & 0 & 129 & 0 & 697 & 496 \\ \hline
\end{tabular}
\end{adjustbox}
\caption{\textbf{Impact of Mu Parameter Changes for S8P with Baseline = 0.67}}
\label{table2}
\end{table}

Row 6 of Table \ref{table2} shows the baseline output where ER1 and ER2 are the same and all clusters and unchanged (UC = 255). Not surprisingly, decreasing mu results in fewer clusters, and increasing mu results in more clusters. However, Table \ref{table2} does show a couple of interesting insights.

First, Table \ref{table2} shows that for S8P, the impact of increasing and decreasing mu is predictable. As noted, increases in mu only caused ER1 clusters to be partitioned, and decreases in mu only cause ER1 clusters to merge. None of the mu changes result in overlapping with ER2 clusters.

The second observation from Table \ref{table2} is that for S8P, clustering is somewhat sensitive to changes in mu. Simply increasing mu from 0.67 to 0.77 dramatically increases the number of clusters from 255 to 451. A 15\% increase in mu resulted in 41\% of the ER1 cluster to be partitioned and a 176\% increase in the overall cluster count.  This suggests that if one believes that mu = 0.67 is near the value for the best DWM clustering results for S8P given the other parameter settings, then the analysis should focus on the impact mu changes around 0.67 in smaller increments, such as 0.01 instead of 0.10.

Again, it is important to note that when working without a truth set, the CCMS is only showing the changes. Whether the changes are resulting in better or worse outcomes requires further analysis. But again, the CCMS can help in this regard as shown in the second example.

Table \ref{table3} shows the effects of changing the Epsilon parameter value. Epsilon is the minimum quality measure for keeping a cluster. The quality of a cluster is measured by a modified Shannon entropy calculation. Clusters that fail to meet the Epsilon threshold are not kept and their references are returned to the input for re-washing in the next iteration after Mu has been incremented. Washing cycles continue until all clusters have a quality score above Epsilon or Mu reaches 1.00. For the S8P dataset, the optimal Epsilon is quite small and lowering it by 0.05 does not change the result. On the other hand, larger values of Epsilon impose higher levels of cluster quality and results in smaller clusters as shown by the increasing number of Partitioned cases and more singleton clusters.

\begin{table}[ht]
\centering
\setlength{\tabcolsep}{6pt}
\renewcommand\arraystretch{1.2}
\begin{adjustbox}{max width=\linewidth}
\begin{tabular}{|c|c|c|c|c|c|c|c|c|}
\hline
\textbf{Epsilon} & \textbf{ER1 Clus} & \textbf{ER1 Sing} & \textbf{Unchanged} & \textbf{Merged} & \textbf{Partitioned} & \textbf{Overlapping} & \textbf{ER2 Clus} & \textbf{ER2 Sing} \\ \hline
0.10 & 255 & 94 & 255 & 0 & 0 & 0 & 255 & 94 \\ \hline
\rowcolor{yellow!50}
0.15 & 255 & 94 & 255 & 0 & 0 & 0 & 255 & 94 \\ \hline
0.20 & 255 & 94 & 253 & 0 & 2 & 0 & 257 & 98 \\ \hline
0.25 & 255 & 94 & 250 & 0 & 5 & 0 & 261 & 105 \\ \hline
0.30 & 255 & 94 & 244 & 0 & 11 & 0 & 272 & 121 \\ \hline
0.35 & 255 & 94 & 230 & 0 & 25 & 0 & 305 & 158 \\ \hline
0.40 & 255 & 94 & 219 & 0 & 36 & 0 & 330 & 181 \\ \hline
0.45 & 255 & 94 & 193 & 0 & 62 & 0 & 414 & 250 \\ \hline
0.50 & 255 & 94 & 164 & 0 & 91 & 0 & 506 & 231 \\ \hline
0.55 & 255 & 94 & 141 & 0 & 114 & 0 & 570 & 366 \\ \hline
0.60 & 255 & 94 & 126 & 0 & 129 & 0 & 614 & 397 \\ \hline
\end{tabular}
\end{adjustbox}
\caption{\textbf{Epsilon Parameter Changes (Optimal = 0.15)}}
\label{table3}
\end{table}

Table \ref{table4} shows the effects of changing the Beta parameter value. Beta controls the DWM blocking process. It represents the minimum frequency of a token that can be used as a blocking token. Blocks are formed by collecting references that share the same blocking token (or alternatively, the same pair of blocking tokens). For the S8P dataset, the Beta value has low sensitivity. Increasing its value up to 31 has no effect on the resulting clusters but will slow down the processing because it will create more blocking tokens and consequently, more blocks to process. Lowering Beta to 17 only causes one ER1 cluster to partition into two ER2 clusters, one of which is a singleton cluster. A wider range of Beta values would likely show additional changes.

\begin{table}[ht]
\centering
\setlength{\tabcolsep}{6pt}
\renewcommand\arraystretch{1.2}
\begin{adjustbox}{max width=\linewidth}
\begin{tabular}{|c|c|c|c|c|c|c|c|c|c|}
\hline
\textbf{Beta} & \textbf{F-Meas} & \textbf{ER1 Clusters} & \textbf{ER1 Single} & \textbf{Unchanged} & \textbf{Merged} & \textbf{Partitioned} & \textbf{Overlapping} & \textbf{ER2 Clusters} & \textbf{ER2 Single} \\ \hline
17 & 0.8079 & 255 & 94 & 254 & 0 & 1 & 0 & 256 & 95 \\ \hline
19 & 0.8079 & 255 & 94 & 254 & 0 & 1 & 0 & 256 & 95 \\ \hline
21 & 0.8079 & 255 & 94 & 254 & 0 & 1 & 0 & 256 & 95 \\ \hline
\rowcolor{yellow!50}
23 & 0.8082 & 255 & 94 & 255 & 0 & 0 & 0 & 255 & 94 \\ \hline
25 & 0.8082 & 255 & 94 & 255 & 0 & 0 & 0 & 255 & 94 \\ \hline
27 & 0.8082 & 255 & 94 & 255 & 0 & 0 & 0 & 255 & 94 \\ \hline
29 & 0.8082 & 255 & 94 & 255 & 0 & 0 & 0 & 255 & 94 \\ \hline
31 & 0.8082 & 255 & 94 & 255 & 0 & 0 & 0 & 255 & 94 \\ \hline
\end{tabular}
\end{adjustbox}
\caption{\textbf{Beta Parameter Changes (Optimal = 23)}}
\label{table4}
\end{table}

\subsection{Industry Research: Materials Classification}

The second example compares two different ER systems in an industry application. The first ER system is the Data Washing Machine (DWM) as described earlier. The second ER system is an adaptation of the DWM by the PiLog Group \citep{PiLog2024} research and development team (denoted by PDWM) to help with classifying engineering parts and materials.

While the PiLog team conducted extensive research and testing, only the results for two datasets is described here to illustrate the utility of CCMS. The two datasets, Testset 100 and Testset 5000 with 100 and 5,000 records, respectively, contain references to parts and materials with item descriptions such as motors and ball bearings, feature descriptions such as voltage, sizes, color, and other technical details. The goal of the PiLog research was to determine if adding their specific knowledge (labeling) of material descriptions to the PDWM would enhance its performance in comparison to the generic, open-source version of the DWM, and potentially make the PDWM a useful tool for their internal processing.

Through these experiments, our objective was to conduct a comprehensive cluster comparison, distinguished by its four distinct case counts, providing a detailed insight into how clusters reconfigure or maintain stability, thereby offering valuable insights into the nature of changes or consistencies between two clustering results.

Our approach involved organizing the dataset and conducting meticulous sorting based on the record identifier (RecID) and cluster IDs generated by ER1 and ER2. This structured sorting laid the foundation for a comprehensive analysis wherein we categorized clusters into distinct cases—Unchanged, Merged, Partitioned, and Overlapping. 

In addition to the cluster comparison, we will be creating a detailed report on the case counts, total ER1s, total ER2s, and singletons. This comprehensive report will provide a thorough understanding of the clustering outcomes and performance metrics of each data washing machine, offering valuable guidance for data management strategies in similar contexts. 

By scrutinizing these clusters, we gained valuable insights into how the two washing machines reconfigure or maintain cluster stability, thereby offering valuable insights into the nature of 
Data Set Condition changes or consistencies between the clustering results.

Additionally, our experiments extended to evaluating auxiliary metrics such as the total number of clusters identified by ER1 and ER2, as well as the count of singleton clusters within both outputs. These auxiliary metrics provided further insights into the clustering behavior and performance of each system. Through our experiments and analysis, we aimed to shed light on the capabilities and limitations of each data washing machine, offering valuable guidance for data management strategies in similar contexts.

We conducted experiments on two datasets as shown in Table \ref{table5}, Test set 100 and Test set 5000, to evaluate the performance of the University Data Washing Machine (UALR DWM) compared to both Classification and Non-Classification variants of the Pilog Data Washing Machine (Pilog DWM). The results demonstrated notable differences in how these systems handled entity resolution, depending on the dataset and the variant used.

\begin{table}[ht]
\centering
\renewcommand\arraystretch{1.2}
\setlength{\tabcolsep}{6pt}
\begin{adjustbox}{max width=\linewidth}
\begin{tabular}{|l|c|c|c|c|}
\hline
\multirow{2}{*}{\textbf{Metric}} & \multicolumn{2}{c|}{\textbf{Classification}} & \multicolumn{2}{c|}{\textbf{Non-Classification}} \\ \cline{2-5}
 & \textbf{Test set 100} & \textbf{Test set 5000} & \textbf{Test set 100} & \textbf{Test set 5000} \\ \hline
ER1 Clusters & 80 & 3918 & 80 & 3918 \\ \hline
ER1 Singletons & 68 & 3423 & 68 & 3423 \\ \hline
Unchanged & 65 & 2517 & 45 & 2464 \\ \hline
Merged & 10 & 918 & 23 & 993 \\ \hline
Partitioned & 5 & 236 & 6 & 334 \\ \hline
Overlapping & 0 & 247 & 6 & 127 \\ \hline
ER2 Clusters & 81 & 4205 & 80 & 4207 \\ \hline
ER2 Singletons & 67 & 3621 & 66 & 3625 \\ \hline
\end{tabular}
\end{adjustbox}
\caption{\textbf{Results of Case Counts with respect to UALR DWM and PiLog DWM under Classification and Non-Classification Conditions}}
\label{table5}
\end{table}

For Test set 100, the UALR DWM identified a total of 80 clusters with 68 singletons, showing consistent results across both comparisons. However, the Pilog DWM exhibited distinct cluster transformation behaviors between its Classification and Non-Classification variants. When compared to the Classification Pilog DWM, 65 clusters remained unchanged, 10 merged into larger clusters, and 5 were partitioned into smaller clusters, while ER2 identified 81 clusters with 67 singletons. On the other hand, the Non-Classification Pilog DWM showed more variation, with 45 unchanged clusters, 23 merged clusters, 6 partitioned clusters, and 6 overlapping clusters, resulting in 80 clusters with 66 singletons in ER2.

The results for Test set 5000 further highlighted these differences. The UALR DWM identified 3,918 clusters with 3,423 singletons, a result consistent across comparisons. However, the Classification Pilog DWM revealed significant changes: 2,517 clusters remained unchanged, 918 merged, 236 were partitioned, and 247 overlapped, with ER2 identifying 4,205 clusters and 3,621 singletons. In contrast, the Non-Classification Pilog DWM produced slightly different outcomes, with 2,464 unchanged clusters, 993 merged, 334 partitioned, and 127 overlapping clusters, leading to 4,207 clusters and 3,625 singletons in ER2.

These experiments underscore how the performance of data washing machines can vary significantly depending on the dataset and the presence of classification capabilities. For smaller datasets like Test set 100, the UALR DWM produced stable cluster counts, but the Pilog DWM's transformations differed substantially between its variants. For larger datasets like Test set 5000, the differences became even more pronounced, reflecting the impact of classification on entity resolution outcomes. The results also revealed the complexity of entity resolution tasks, especially with factors such as merged, partitioned, and overlapping clusters. These findings emphasize the need for tailored strategies for different datasets and demonstrate how the proposed metric can help identify areas for optimization and improvement in entity resolution systems.

\section{Conclusion}

The Parameter Sensitivity and Materials Classification examples presented here demonstrate that the Case Count Metric System (CCMS) can be useful tool for comparing cluster ER outcomes in situations where a truth set is not available to automatically compute and compare precision, recall, F-measure and other tradition ER metrics. It also is a significant improvement over the single-valued TWI metric by giving a broader insight into the nature of the cluster changes and its ability to display and analyze example differences. 

By breaking down cluster transformations into the four mutually exclusive categories of Unchanged, Merged, Partitioned, and Overlapping, CCMS provides a deeper and more actionable understanding of how clusters are transformed in reaction to changes within or between ER systems. This type of granular analysis helps to identify whether specific changes to a cluster ER system are net positive or net negative, making it a useful tool for fine-tuning performance in complex data environments. 

While CCMS has proved to be robust and insightful in a research setting, its performance in noisier, highly heterogeneous, and large-scale datasets requires further investigation. The application of CCMS in a wider variety of domains such as healthcare, finance, and supply chain management need further investigation and validation.

\section*{Author Contributions}

JRT conceptualized the study, developed the main methodology, and led the writing of the manuscript. MAM contributed to the data analysis and validation processes. MCC assisted with manuscript reviewing, editing, and formatting for publication. MKM, OKM, and KS participated in team discussions, provided critical feedback, and supported refinement of the experimental design and interpretation of results. LC served as the PiLog data representative and coordinated access to the datasets used in this research. All authors reviewed and approved the final manuscript and agree to be accountable for the work presented herein.

\section*{Funding}
This research was partially supported by the National Science Foundation under EPSCoR Award No. OIA-1946391 and the PiLog Group.

\bibliographystyle{Frontiers-Harvard}
\bibliography{test}

\end{document}